\bigskip
\documentclass[twocolumn,preprintnumbers,amsmath,amssymb]{revtex4}
\usepackage{graphicx}
\usepackage{fancyhdr}
\usepackage{dcolumn}
\usepackage{bm}
\usepackage{float}
\usepackage{graphicx}
\usepackage{multirow}
\usepackage{amssymb}
\usepackage{amsmath}
\usepackage{graphicx}
\usepackage{float}
\usepackage[normalem]{ulem}
\usepackage[dvips]{color}

\begin{document}
\preprint{ }
\title{ Pion production in intermediate-energy heavy-ion collisions with a relativistic quantum molecular dynamics model }
\author{ Si-Na Wei}
\email{Electronic address: 471272396@qq.com}
\author{ Zhao-Qing Feng}
\email{Corresponding author: fengzhq@scut.edu.cn}

\affiliation{School of Physics and Optoelectronics, South China University of Technology, Guangzhou 510640, China}
\begin{abstract}
The relativistic mean field approach by distinguishing the slope of symmetry energy is implemented into the Lanzhou quantum molecular dynamics transport model (LQMD.RMF). The collective flows in the isotopic nuclear reactions are systematically investigated by the relativistic quantum molecular dynamics model with various slopes of symmetry energy. The structure of the directed and elliptic flows is consistent with the results of the nonrelativistic transportation of nucleon system. The directed flow difference between free neutrons and protons appears in the midrapidity region. The transverse momentum spectra of $\pi^+$ production is close to each other in the nearly symmetric $^{108}\mathrm{Sn} + ^{112}\mathrm{Sn}$ system and the neutron-rich  $^{132}\mathrm{Sn} + ^{124}\mathrm{Sn}$ system. However, since there are more neutron-neutron scatterings in neutron-rich system, the transverse momentum spectra of $\pi^-$ production in the neutron-rich system are higher than one in the nearly symmetric system. For a given reaction system,  the transverse momentum spectra of $\pi^+$ and $\pi^-$  production are independent on the stiffness of symmetry energy. This leads to the fact that the single ratio and the double ratio  are independent on the stiffness of symmetry energy.  Moreover, the double ratio without the $\pi$-nucleon potential decreases with increasing the  transverse momentum. However,  the double ratio with the inclusion of $\pi$ potential increases with increasing the  transverse momentum.
\end{abstract}

\maketitle

\section{INTRODUCTION}
The equation of state (EOS) of nuclear matter, which originates from the nucleon-nucleon interaction, plays an important role in heavy-ion collisions and  properties of nuclei and neutron star (NS).  To extract the nuclear EOS,  the heavy-ion collisions, properties of nuclei and  NS have been widely studied. Since the nuclear many-body problems are highly nonlinear and the EOS is  not a direct observable quantity in experiments, there are still some uncertainties  in the EOS after great effort\cite{rmq1,rmq2,rmq3,rmqa1,rmqa2,rmqa3}. For instance,   the EOS of nuclear matter extracted from the data of the heavy-ion collisions still has  uncertainties at high nuclear density\cite{rmq1}, and the EOS extracted from GW170817 event also has uncertainties at high nuclear density\cite{rmq2}. Although the EOS can be extracted from properties of NS, the internal composition of NS is still poorly understood. The core of  NS may  contain  exotic materials,  such as hyperons,  kaons, pions and deconfined quark matter\cite{rmq4,rmq5,rmq6,rmq7,rmq8}. The heavy-ion collisions in terrestrial laboratory provide a unique possibility to study both the EOS and  exotic materials.

The collective flows of  heavy-ion collision were proposed  in the 1970's  and  first detected  in experiment at Bevalac\cite{rmq9,rmq10,rmq11,rmq12}. Since the  collective flows are associated with the nucleon-nucleon interaction, nucleon-nucleon scattering etc, the collective flows have been used to extract the nuclear EOS. The  collective flows are also helpful in understanding the phase transition between hadronic matter and quark matter. The  collective flows after hadron-quark phase transition would have some differences compared to the  collective flows of pure hadronic matter. Generally, when the  phase transition between hadronic matter and quark matter  occurs,  the collective flows  of  heavy-ion collision indicate a soft EOS \cite{rmq13,rmq14,rmq15,rmq16}.
Besides, the ratios of isospin particles in heavy-ion collisions, such as $\pi^-/\pi^+$, $K^0/K^+$ and $\Sigma^-/\Sigma^+$, are though to be sensitive to  the isospin asymmetric part of EOS (the nuclear symmetry energy)\cite{rmq17,rmq18,rmq19,rmq20,rmq21,rmq22,rmq23}. In experiment, the production of pions and kaons has been measured in $\rm^{197}Au$+$\rm^{197}Au$ collisions. The $K^+$ production predicted by various transport model favors a soft EOS at high baryon densities\cite{rmq24,rmq25,rmq26,rmq27,rmq28}. However,  the $\pi^-/\pi^+$ ratio predicted by various transport models is still model dependent\cite{rmq29,rmq30,rmq31,rmq32}.  Based on the FOPI data of  the $\pi^-/\pi^+$ ratio\cite{rmq33}, some results  favor a stiff  symmetry energy\cite{rmq29,rmq30}, however, others conclusions imply a soft symmetry energy\cite{rmq31,rmq32}. Recently, by analysing the ratios of the charged pion in $\rm^{132}Sn + ^{124}Sn$, $\rm^{112}Sn +^{124}Sn$ and $\rm^{108}Sn + ^{112}Sn$ collisions\cite{rmq34},   the slope of the symmetry energy ranging from  42 to 117 MeV was predicted\cite{rmq35}.  In different transport models, the relation between the ratios of the charged pion and the properties of symmetry energy may be different. It is still worthwhile to study the ratios of the charged pion in different transport models.

As one of the popular  transport models, the quantum molecular dynamics (QMD) approach has been developed into many versions and has been used to describe the heavy-ion collisions successfully.  At high-energy heavy-ion collisions, since the  relativistic effects become significant, the relativistic effects should be taken account in the QMD. The  relativistic QMD (RQMD)  approach was proposed in this purpose\cite{rmq36,rmq37}. Recently, the  relativistic mean meson field theory has been implemented into the QMD model (QMD.RMF)\cite{rmq38,rmq39,rmq40}. The RQMD.RMF has been applied to investigate the collective flows of hadrons successfully\cite{rmq38,rmq39,rmq40}.  In this work, we implement the relativistic mean-field theory with isovector-vector and isovector-scalar fields into the Lanzhou quantum molecular dynamics  model  (LQMD.RMF) to investigate the  collective flows and the ratios of the charged pion. The  channel of generation and decay of resonances ($\Delta$(1232), N*(1440), N*(1535), etc), hyperons and mesons has been included\cite{rmq28,rmq29,rmqa4} in the LQMD model. With the LQMD.RMF, we are able to explore the relation between  the EOS and and the properties of the charged pion.

The paper is organized as follows. In Sec. II, we briefly introduce formulas and approaches used in this work. The formulas include RMF theory,  dispersion relation and production of pion. Results and discussions are presented in Sec. III. A summary is finally given in Sec. IV.

\section{Formalism}
\subsection{ Relativistic mean field theory}
The interaction of RMF is achieved by exchanging mesons. The scalar and vector mesons provide the medium-range attraction and short-range repulsion between the nucleons, respectively\cite{rmq41}.  The nonlinear self-interaction of the $\sigma$ meson is introduced to bring down the incompressibility to a reasonable domain\cite{rmq42}. To investigate the properties of symmetry energy, we also consider the  isovector-vector $\rho$ \cite{rmq43} and the isovector-scalar $\delta$ mesons\cite{rmq44}. The Lagrangian density is written as
\begin{eqnarray}
&\mathcal{L}=&\bar{\psi}[\gamma_{\mu}(i\partial^{\mu}-g_{\omega}{\omega}^{\mu}-g_{\rho}\vec{\tau}\cdot \vec{b}^{\mu})-(M_N-g_{\sigma}\sigma\nonumber\\&&-g_\delta\vec{\tau}\cdot\vec{\delta})]\psi
+\frac{1}{2}(\partial_{\mu}\sigma\partial^{\mu}\sigma-m_{\sigma}^2\sigma^2)-\frac{1}{3}g_2\sigma^3\nonumber\\
&&-\frac{1}{4}g_3\sigma^4
+\frac{1}{2}m_{\omega}^2{\omega}_{\mu}{\omega}^{\mu}-\frac{1}{4}F_{\mu\nu}F^{\mu\nu}+\frac{1}{2}m_{\rho}^2\vec{b}_{\mu}\vec{b}^{\mu}\nonumber\\&&
-\frac{1}{4}\vec{B}_{\mu\nu}\vec{B}^{\mu\nu}+\frac{1}{2}(\partial_{\mu}\vec{\delta}\cdot\partial^{\mu}\vec{\delta}-m_{\delta}^2\vec{\delta}^2),
\label{pos1}
\end{eqnarray}
where $M_N=938$ MeV is the nucleon mass in the free space. $g_i$ with $i=\sigma, \omega, \rho, \delta$ is the coupling constants between the nucleon. $m_i$ with $i=\sigma, \omega, \rho, \delta$ is the meson masses. $g_2$ and $g_3$ are the coupling constants of the nonlinear self-interaction of $\sigma$ meson. $F_{\mu\nu}=\partial_{\mu}\omega_{\nu}-\partial_{\nu}\omega_{\mu}$ and $\vec{B}_{\mu\nu}=\partial_{\mu}\vec{b}_{\nu}-\partial_{\nu}\vec{b}_{\mu}$ are the strength tensors of $\omega$ and $\rho$ mesons, respectively. The equations of motion for the nucleon and meson are obtained from the Euler-Lagrange equations, and written as
\begin{equation}
[i\gamma^{\mu}\partial_{\mu}-g_{\omega}\gamma^{0} \omega_{0}-g_{\rho}\gamma^{0}b_{0}\tau_3-(M_N-g_{\sigma}\sigma-g_\delta\tau_3\delta_3)]\psi=0
\label{qen1}
\end{equation}
\begin{equation}
m_{\sigma}^2\sigma+g_2\sigma^2+g_3\sigma^3=g_\sigma\bar{\psi}\psi=g_{\sigma}\rho_S
\label{qen2}
\end{equation}
\begin{equation}
m_{\omega}^2 \omega_{0}=g_\omega\bar{\psi}\gamma^0\psi=g_{\omega} \rho
\label{qen3}
\end{equation}
\begin{equation}
m_{\rho}^2 b_0=g_\rho\bar{\psi}\gamma^0\tau_3\psi=g_{\rho} \rho_{3}
\label{qen4},
\end{equation}
\begin{equation}
m_{\delta}^2 \delta_3=g_\delta\bar{\psi}\tau_3\psi=g_{\delta} \rho_{S3}
\label{qen5},
\end{equation}
where $\rho$ and $\rho_S$ are  the baryon and the scalar densities, respectively. $\rho_{3} =\rho_{p}-\rho_{n}$  is the difference between the proton and neutron densities, and $\rho_{S3} =\rho_{Sp}-\rho_{Sn}$  is the difference between the proton and neutron scalar densities.

\begin{figure}
\includegraphics[height=6cm,width=8cm]{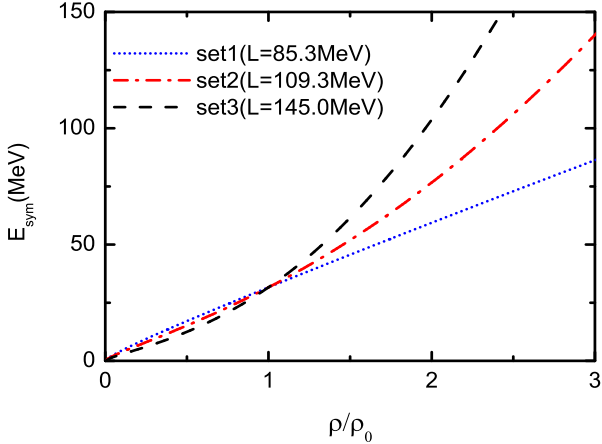}
\caption{(Color online) The symmetry energy  as a function of the baryon density.}\label{esym}
\end{figure}

In the RMF approximation, the energy density  is given as
\begin{eqnarray}
\epsilon&=&\sum_{i=n,p}2\int_0^{p_F}\frac{d^3p}{(2\pi)^3}\sqrt{p^2+M_i^2}
+\frac{1}{2}m_{\sigma}^2\sigma^2+\frac{1}{3}g_2\sigma^3   \nonumber\\
&&+\frac{1}{4}g_3\sigma^4
+\frac{1}{2}m_\omega^2\omega_0^2+\frac{1}{2}m_\rho^2b_0^2+\frac{1}{2}m_\delta^2\delta_3^2
\label{en1},
\end{eqnarray}
where $p_F$ is the nucleon Fermi momentum. $M_i=M_N-g_\sigma\sigma\mp g_\delta\delta_3$ ($-$ proton, $+$ neutron) is the effective nucleon mass. With the isospin asymmetry parameter $\alpha=(\rho_{n}-\rho_{p})/(\rho_{n}+\rho_{p})$, the symmetry energy is written as\cite{rmq44}
 \begin{eqnarray}
E_{sym}&&=\frac{1}{2}\frac{\partial^2E(\rho,\alpha)}{\partial\alpha^2}|_{\alpha=0}       \nonumber\\
&&=\frac{1}{6}\frac{p_F^2}{E_F^*}+\frac{1}{2}f_\rho\rho
-\frac{f_\delta}{2}\frac{M^{*2}\rho}{E_F^{*2}[1+f_\delta A(p_F,M^*)]}
\label{esy1},
\end{eqnarray}
where $f_i\equiv\frac{g_i^2}{m_i^2}$, $i=\rho,\delta$. $E_F^*=\sqrt{p_F^2+M^{*2}}$ and $M^*=M_N-g_\sigma\sigma$  is the effective nucleon mass of symmetric nuclear matter. The integral $A(p_F,M^*)$ is defined as
 \begin{eqnarray}
A(p_F,M^*)&&=\frac{4}{(2\pi)^3}\int d^3p\frac{p^2}{(p^2+M^{*2})^{3/2}}\nonumber\\&&=3(\frac{\rho_S}{M^*}-\frac{\rho}{E_F^*})
\label{esy2}.
\end{eqnarray}
In this work, we set the saturation density as $\rho_0=0.16 fm^{-3}$. The  binding energy per particle of symmetry nuclear matter is set to be $E/A-M_N=-16$ MeV. For symmetric nuclear matter, we set set1, set2 and set3 models to be the same  as a result of vanishing isospin asymmetry. As shown in  Table \ref{tb1} and Fig.\ref{esym}, the symmetry energy of set1, set2 and set3 is set to be 31.6 MeV at saturation density. The set1 contains only $\rho$  meson,  however, set2 and  set3 contains both the  $\rho$ and $\delta$ mesons. For set1, when the symmetry energy is set to be  31.6 MeV at saturation density, the coupling parameters $g_\rho$ is fixed, and the slope of symmetry is fixed to $L=85.3$ MeV.
For set2 and set3, the slope of symmetry energy is obtained by varying  the coupling parameters $g_\rho$ and $g_\delta$.  The symmetry energy with both the  $\rho$ and $\delta$ mesons can not be softer than one of set1 containing only $\rho$ meson. In order to broad the range of the slope parameter, we set the slope parameter of set2 and set3 to be 109.3 and 145.0 MeV by varying  the coupling parameters $g_\rho$ and $g_\delta$, respectively. A broader range of the slope parameter would be helpful to understand the relation between  the properties of symmetry energy and the observables of heavy-ion collisions.

\begin{table*}[!htp]
\begin{center}
\caption{ Parameter sets for RMF. The saturation density $\rho_0$ is set to be 0.16 $fm^{-3}$. The
binding energy at the saturation density is $E/A-M_N=-16$ MeV.  The isoscalar-vector $\omega$ and isovector-vector $\rho$ masses are fixed to their physical values, $m_{\omega}$ = 783 MeV and $m_{\rho}$ = 763 MeV. The remaining meson mass $m_\sigma$ is set to be 550 MeV. \label{tb1}}
\footnotesize
\begin{tabular*}{175mm}{c|@{\extracolsep{\fill}}c|c|c|c|c|c|c|c|c|c|c|c|c}
\hline
  model&   $g_\sigma$  &    $g_\omega$ &   $g_2$ ($fm^{-1}$) &   $g_3$ &   $g_\rho$  &   $g_\delta$ &  $K$ (MeV) & $E_{sym}$($\rho_0$) (MeV) & $L$ ($\rho_0$)(MeV) \\
set1  & 8.145 & 7.570 &  31.820  & 28.100 &4.049&-& 230& 31.6 &85.3 \\
set2  & 8.145 & 7.570  & 31.820  & 28.100 &8.673&5.347& 230&31.6&109.3  \\
set3  & 8.145 &  7.570 & 31.820  &28.100 & 11.768 &     7.752   &  230  &31.6&145.0 \\
\hline
\end{tabular*}
\end{center}
\end{table*}
\subsection{Relativistic quantum molecular dynamics approach}
In order to ivestigate high-energy heavy-ion collision, the RQMD was proposed\cite{rmq36,rmq37}. Recently,  the RMF has been implemented into the RQMD\cite{rmq38,rmq39,rmq40}. In RQMD, for  $N$-body system, there are 4$N$ position coordinates $q_i^\mu$ and 4$N$ momentum coordinates $p_i^\mu$ ($i=1,...,N$).  However, the physical trajectories ($\vec{q}_i$ and $\vec{p}_i$) are
6$N$ for  $N$-body system. 2$N$ constraints are needed to  reduce the number of dimensions from 8$N$ to  physical trajectories 6$N$\cite{rmq36,rmq37,rmq38,rmq39,rmq40,rmq45,rmq46,rmqa6},
\begin{eqnarray}
\phi_i\approx 0 (i=1,...,2N)
\label{rqmd1},
\end{eqnarray}
where 2$N$ constraints  satisfy  the physical 6$N$ phase space. The sign $\approx$ means Dirac's weak equality. The on-mass shell conditions is able to reduces the phase space from 8$N$  to 7$N$ dimensions,
\begin{eqnarray}
\phi_i\equiv p_i^{*2}-M_i^{*2}=(p_i-V_i)^2-(M_N-S_i)^2=0
\label{rqmd2},
\end{eqnarray}
here $i=1,...,N$. The remaining $N$ constraints are the time  fixation constraints. A  simple choice of the time  fixation constraints, which obey the world-line condition, are written as\cite{rmq37,rmq45,rmqa6,rmqa5}
\begin{eqnarray}
&&\phi_{i+N}\equiv\hat{a}\cdot(q_i-q_N)=0, (i=1,...,N-1), \nonumber\\&&
\phi_{2N}\equiv\hat{a}\cdot q_N-\tau=0,
\label{rqmd3}
\end{eqnarray}
where $\hat{a}=(1,\vec{0})$  is the four-dimensional unit-vector\cite{rmq36,rmq37,rmq38,rmq39,rmq40,rmq45}. In two-body center-of-mass system, $\hat{a}$ is defined as $p_{ij}^\mu/\sqrt{p_{ij}^2}$ with $p_{ij}^\mu=p_i^\mu+p_j^\mu$. We see that only the constraint $i=2N$ depends on $\tau$. With the above 2 $N$ constraints, the number of dimensions 8$N$ will reduce to 6$N$. These 2$N$ constraints are  conserved in time:
\begin{eqnarray}
\frac{d\phi_i}{d\tau}&&=\frac{\partial \phi_i}{\partial\tau}+\sum_k^{2N}\lambda_k[\phi_i,\phi_k]=0\nonumber\\&&
=\frac{\partial \phi_i}{\partial\tau}+\sum_k^{2N}C_{ik}^{-1}\lambda_k=0.
\label{rqmda1}
\end{eqnarray}
Since only the constraint $i=2N$ depends on $\tau$,  $\lambda$ is written as\cite{rmq45}
\begin{eqnarray}
\lambda_i=-C_{2N,i}\frac{\partial\phi_{2N}}{\partial\tau}, (i = 1,...,2N-1),
\label{rqmd5}
\end{eqnarray}
with $C_{ij}^{-1}=[\phi_i,\phi_j]$. The Poisson brackets are defined as
\begin{eqnarray}
[A,B]=\sum_k(\frac{\partial A}{\partial p_k}\cdot\frac{\partial A}{\partial q_k}-\frac{\partial B}{\partial q_k}\cdot\frac{\partial B}{\partial p_k}).
\label{rqmd6}
\end{eqnarray}
Follow previous studies, the Hamiltonian of the $N$-body system is  constructed as the linear combination of $2N-1$ constraints\cite{rmq45,rmqa6,rmqa5}:
\begin{eqnarray}
H=\sum_{i=1}^{2N-1}\lambda_i(\tau)\phi_i,
\label{rqmd4}
\end{eqnarray}
Assuming  $[\phi_i,\phi_j]=0$,  the $\lambda_i=0$ for $N+1<i<2N$\cite{rmq45}.  The equations of motion are then obtained as
\begin{eqnarray}
&&\frac{dq_i}{d\tau}=[H,q_i]=\sum_j^{N}\lambda_j\frac{\partial\phi_j}{\partial p_i},\nonumber\\&&
\frac{dp_i}{d\tau}=[H,p_i]=-\sum_j^{N}\lambda_j\frac{\partial\phi_j}{\partial q_i},
\label{rqmd7}
\end{eqnarray}
with the  on-mass shell conditions (Eq.(\ref{rqmd2})) as inputs,  the equations of motion are obtained  as
\begin{eqnarray}
&&\dot{\vec{r}}_i=\frac{\vec{p}_i^*}{p_i^{*0}}+\sum_{j=1}^N(\frac{M_j^*}{p_j^{*0}}\frac{\partial M_j^*}{\partial \vec{p_i}}+z_j^{*\mu}\cdot\frac{\partial V_{j\mu}}{\partial\vec{p_i}}),\nonumber\\&&
\dot{\vec{p}}_i=-\sum_{j=1}^N(\frac{M_j^*}{p_j^{*0}}\frac{\partial M_j^*}{\partial \vec{r_i}}+z_j^{*\mu}\cdot\frac{\partial V_{j\mu}}{\partial\vec{r_i}}),
\label{rqmd8}
\end{eqnarray}
where $z_i^{*\mu}=p_i^{*\mu}/p_i^{*0}$ and $M_i^*=M_N-S_i$.  The scalar potential $S_i$ and the vector potential $V_{i\mu}$ in RQMD are written as
\begin{eqnarray}
&&S_i=\frac{1}{2}g_\sigma\sigma_i+\frac{1}{2}g_\delta t_i\delta_i\nonumber\\&&
V_{i,\mu}=\frac{B_i}{2}g_\omega\omega_{i,\mu}+\frac{B_it_i}{2}g_\rho b_{i,\mu}
\label{rqmd9}
\end{eqnarray}
here $t_i=1$ for  protons and  $t_i=-1$  for neutrons.  $B_i$ is the baryon number of the $i$th particle. The meson field is obtained from RMF:
\begin{eqnarray}
&&m_\sigma^2\sigma_i+g_2\sigma_i^2+g_3\sigma_i^3=g_\sigma\rho_{S,i},\nonumber\\&&
m_\omega^2\omega_i^\mu=g_\omega J_i^\mu,\nonumber\\&&
m_\delta^2\delta_i=g_\delta(\rho_{Sp,i}-\rho_{Sn,i})=g_\delta\rho_{S3,i},\nonumber\\&&
m_\rho^2b_i=g_\rho(\rho_p-\rho_n)=g_\rho R_i^\mu.
\label{rqmd10}
\end{eqnarray}
In the RQMD approach, the scalar density, the isovector-scalar density,  the baryon current and the isovetor baryon current are written as
\begin{eqnarray}
&&\rho_{S,i}=\sum_{j\neq i}\frac{M_j}{p_j^0}\rho_{ij},\qquad \rho_{S3,i}=\sum_{j\neq i}t_j\frac{M_j}{p_j^0}\rho_{ij},\nonumber\\&&
J_i^\mu=\sum_{j\neq i}B_j\frac{p_j^\mu}{p_j^0}\rho_{ij},\quad R_i^\mu=\sum_{j\neq i}t_jB_j\frac{p_j^\mu}{p_j^0}\rho_{ij}.
\label{rqmd11}
\end{eqnarray}
Since  the difference between the numerical results  by using effective mass $M_j^*$ and kinetic momentum $p_j^{\mu*}$ in density and current and those by using  a free mass $M_j=M_N=938$ MeV and canonical momentum $p_j^{\mu}$ in density and current is small, a free mass $M_j=M_N=938$ MeV and canonical momentum $p_j^{\mu}$ have been used in the above density and current\cite{rmq39}.  The interaction density $\rho_{ij}$ is given by the Gaussian:
\begin{eqnarray}
&&\rho_{ij}=\frac{\gamma_{ij}}{(4\pi L)^{3/2}}\mathrm{exp}(\frac{q_{T,ij}^2}{4L}),
\label{rqmd12}
\end{eqnarray}
where $q_{T,ij}^2$ is  a distance squared. $\gamma_{ij}$  is a Lorentz factor  ensuring the correct normalization of the Gaussian\cite{rmq47}, and equals $(p_i^0+p_j^0)/(p_i+p_j)$  in two-body center-of-mass frame. In this work, we set the square of wave-packet width as $L=2.0 {fm^2}$.

\subsection{The dispersion relation and production of pion}
The Hamiltonian of mesons  is defined as\cite{rmqa4,rmq48,rmq49,rmq50}
\begin{eqnarray}
H_M=\sum_{i=1}^{N_M}[V_i^C+\omega(\vec{p}_i,\rho_{i})],
\label{rqmd13}
\end{eqnarray}
$V_i^C$ is the Coulomb potential, and is written as
\begin{eqnarray}
V_i^C=\sum_{j=1}^{N_B}\frac{e_ie_j}{r_{ij}},
\label{rqmd14}
\end{eqnarray}
$N_M$ and $N_B$ are the total numbers of mesons
and baryons including charged resonances, respectively. The pion potential in the medium, which contains the isoscalar and isovector contributions, is defined as
\begin{eqnarray}
\omega_{}(\vec{p}_i,\rho_i)=\omega_{isoscalar}(\vec{p}_i,\rho_{i})+
C_\pi\tau_z\alpha(\rho/\rho_0)^{\gamma_\pi},
\label{rqmd15}
\end{eqnarray}
where  $\alpha$ is the isospin asymmetry parameter. The coefficient $C_\pi$ equals 36 MeV.  The isospin quantity $\tau$ is 1, 0 and -1 for $\pi^{-}$, $\pi^{0}$ and $\pi^{+}$, respectively. $\gamma_\pi$  determines  the isospin splitting of pion potential, and is set to be 2.  In this work, the scalar part of  pion potential $\omega_{isoscalar}$ is chosen as  the $\Delta$-hole model. The  pion potential, which contains
a pion branch (smaller value) and a $\Delta$-hole (larger value) branch, is defined as
\begin{eqnarray}
\omega_{isoscalar}(\vec{p_i},\rho_{i})&&=S_\pi(\vec{p_i},\rho_{i})
\omega_{\pi-like}(\vec{p_i},\rho_{i})+\nonumber\\&&S_\Delta(\vec{p_i},\rho_{i})
\omega_{\Delta-like}(\vec{p_i},\rho_{i}).
\label{rqmd16}
\end{eqnarray}
The probability of  the pion branch and the  $\Delta$-hole branch satisfies the following equation:
\begin{eqnarray}
S_\pi(\vec{p_i},\rho_{i})+S_\Delta(\vec{p_i},\rho_{i})=1,
\label{rqmd17}
\end{eqnarray}
The probability of both the pion branch and the  $\Delta$-hole branch is defined as\cite{rmq50}
\begin{eqnarray}
S(\vec{p_i},\rho_{i})=\frac{1}{1-\partial\Pi(\omega)/\partial\omega^2},
\label{rqmd18}
\end{eqnarray}
where  $\omega$ stands for $\omega_{\pi-like}$ and  $\omega_{\Delta-like}$. The eigenvalues of $\omega_{\pi-like}$ and  $\omega_{\Delta-like}$ are generated from the pion dispersion relation:
\begin{eqnarray}
\omega^2=\vec{p}_i^2+m_\pi^2+\Pi(\omega),
\label{rqmd19}
\end{eqnarray}
where $\Pi$ is the pion self-energy. Including the short-range $\Delta$-hole interaction, the pion self-energy is defined as
\begin{eqnarray}
\Pi=\frac{\vec{p}_i^2\chi}{1-g'\chi},
\label{rqmd20}
\end{eqnarray}
here $m_\pi$ is the pion mass. The Migdal parameter $g'$  is set to be 0.6.  $\chi$ is defined as
\begin{eqnarray}
\chi=-\frac{8}{9}(\frac{f_{\Delta}}{m_\pi})^2
\frac{\omega_\Delta\rho\hbar^3}{\omega_\Delta^2-\omega^2}
\mathrm{exp}(-2\vec{p}_i^2/b^2),
\label{rqmd21}
\end{eqnarray}
where $\omega_\Delta=\sqrt{m_\Delta^2+\vec{p}_i^2}-M_N$, and $m_\Delta$ is the delta masses.  In this work, the
$\pi N\Delta$ coupling constant $f_\Delta$ is 2, and the cutoff factor $b$ is $7m_\pi$.

In this work, we assume that the mass and  energy-momentum of $\Delta$ resonances are not changed by the RMF, and neglect the  threshold effect\cite{rmq56,rmq57,rmq58}. With the energy of pion and  Coulomb potential, the energy balance of this work in the decay of resonances is written as:
\begin{eqnarray}
&&\sqrt{m_R^2+\vec{p}_R^2}=\sqrt{M_N^2+(\vec{p}_R-\vec{p}_\pi)^2}+\omega_\pi(\vec{p}_\pi,\rho)+V_\pi^{C},\nonumber\\&&
\label{rqmdn1}
\end{eqnarray}
where $\vec{p}_R$ and $\vec{p}_\pi$ are the momenta of resonances and pions, respectively. $m_R$ is the mass of resonances.

The pion is generated from the  direct nucleon-nucleon collision and  decay of the resonances $\Delta(1232)$ and $N^* (1440)$. The relation channels of resonances and pions,  which are taken as same as those of LQMD model, are given as follow\cite{rmq28,rmqa4,rmq52,rmq53}:
\begin{eqnarray}
&&NN\leftrightarrow N\Delta,\quad NN\leftrightarrow NN^*, \quad NN\leftrightarrow \Delta\Delta, \quad\Delta\leftrightarrow N\pi, \nonumber\\&& N^*\leftrightarrow N\pi, \quad NN\rightarrow NN\pi(s-state).
\label{rqmdn2}
\end{eqnarray}
For the  production of $\Delta(1232)$ and $N^* (1440)$ resonances in a nucleon-nucleon scattering, the  parameterized cross section calculated by the one-boson exchange model has been employed\cite{rmq54}. The decay width of $\Delta(1232)$ and $N^* (1440)$,  which originates from the p-wave resonances, is momentum-dependent and expressed as\cite{rmq54}
\begin{eqnarray}
\Gamma(|\vec{p}|)=\frac{a_1|\vec{p}|^3}
{(1+a_2|\vec{p}|^2)(a_3+|\vec{p}|^2)}\Gamma_0,
\label{rqmdn3}
\end{eqnarray}
where $|\vec{p}|$ is  the momentum of the created pion. The parameters $a_1$, $a_2$ and $a_3$ are taken as 22.48 (17.22), 39.69 (39.69), 0.04(0.09) for $\Delta (N^*)$, respectively. The bare decay width of $\Delta (N^*)$ is given as $\Gamma_0=0.12 (0.2) \rm{GeV}$. With the momentum-dependent decay width, the cross section of pion-nucleon scattering has the Breit-Wigner form:
\begin{eqnarray}
\sigma_{\pi N}(\sqrt{s})=\sigma_{\rm{max}}(\frac{\vec{p}_0}{\vec{p}})^2
\frac{0.25\Gamma^2(\vec{p})}{0.25\Gamma^2(\vec{p})
+(\sqrt{s}-m_0)^2},
\label{rqmdn4}
\end{eqnarray}
where $\vec{p}$ and $\vec{p}_0$ are the three-momenta of pions at energy of $\sqrt{s}$ and $m_0$, respectively. The  maximum cross section $\sigma_{\rm{max}}$ of $\Delta$ and $N^*$ resonances is obtained by fitting the total cross sections of the experimental data in pion-nucleon scattering  with the Breit-Wigner formula\cite{rmq55}. For instance, the  maximum cross section $\sigma_{\rm{max}}$ of $\Delta$ resonance is  200, 133.33, and 66.7 mb for $\pi^+p\rightarrow \Delta^{++} \quad(\pi^-n\rightarrow \Delta^{-})$, $\pi^0p\rightarrow \Delta^{+} \quad(\pi^0n\rightarrow \Delta^{0})$ and $\pi^-p\rightarrow \Delta^{0} \quad(\pi^+n\rightarrow \Delta^{+})$, respectively\cite{rmq53}.

\section{Results and discussions}
We have implemented the RMF model into the LQMD  model. To check  this model,  we first investigate the collective flows of  $\rm^{108}Sn + ^{112}Sn$ and $\rm^{132}Sn + ^{124}Sn$ collisions. The directed and  elliptic flows come from the Fourier expansion of the azimuthal distribution:
 \begin{eqnarray}
\frac{dN}{d\phi}(y, p_T)&=&N_0[1+2V_1(y, p_T)\mathrm{cos}(\phi)\nonumber\\&&+2V_2(y, p_T)\mathrm{cos}(2\phi)],
\label{rqmd22}
\end{eqnarray}
where the azimuthal angle of the emitted particle $\phi$ is measured from the reaction plane.   $p_T=\sqrt{p_x^2+p_y^2}$ is  the transverse momentum.  The directed flow $V_1$  and elliptic flow $V_2$ are written as
 \begin{eqnarray}
&&V_1\equiv<\mathrm{cos}(\phi)>=<\frac{p_x}{p_T}>, \nonumber\\&& V_2\equiv<\mathrm{cos}(2\phi)>=<\frac{p_x^2-p_y^2}{p_T^2}>,
\label{rqmd23}
\end{eqnarray}
The directed  flow  stands for information on  the azimuthal anisotropy of the transverse emission. The elliptic flow tells us about the competition between the in-plane ($V_2>0$) and out-of-plane $V_2<0$ emissions. Since  interactions of nuclear matter will have impact on collective flows, the collective flows have been used to extract  the high-density behavior of the EOS widely.
\begin{figure}
\includegraphics[height=10cm,width=8cm]{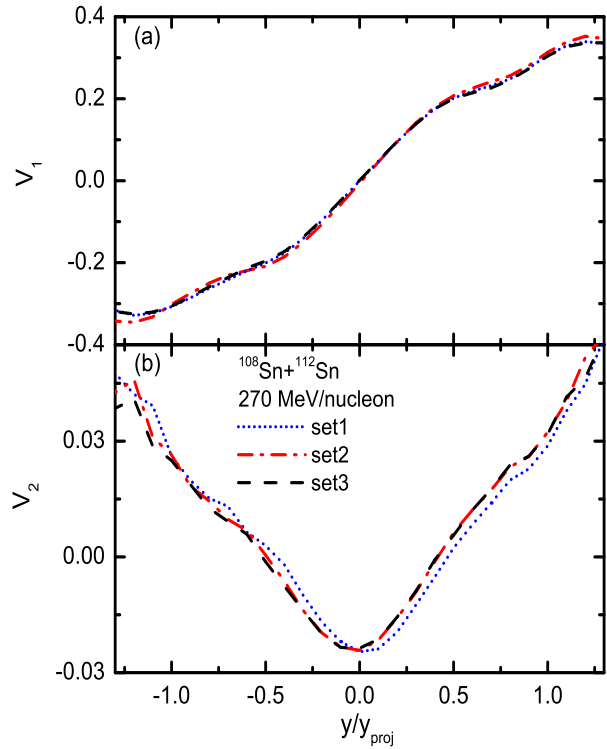}
\caption{(Color online) Rapidity distribution of the collective flows of free neutrons   in the
$^{108}\mathrm{Sn} + ^{112}\mathrm{Sn}$ reaction at an incident energy of 270 MeV/nucleon for impact parameter b=3 fm.}\label{108fl}
\end{figure}
\begin{figure}[H]
\includegraphics[height=10cm,width=8cm]{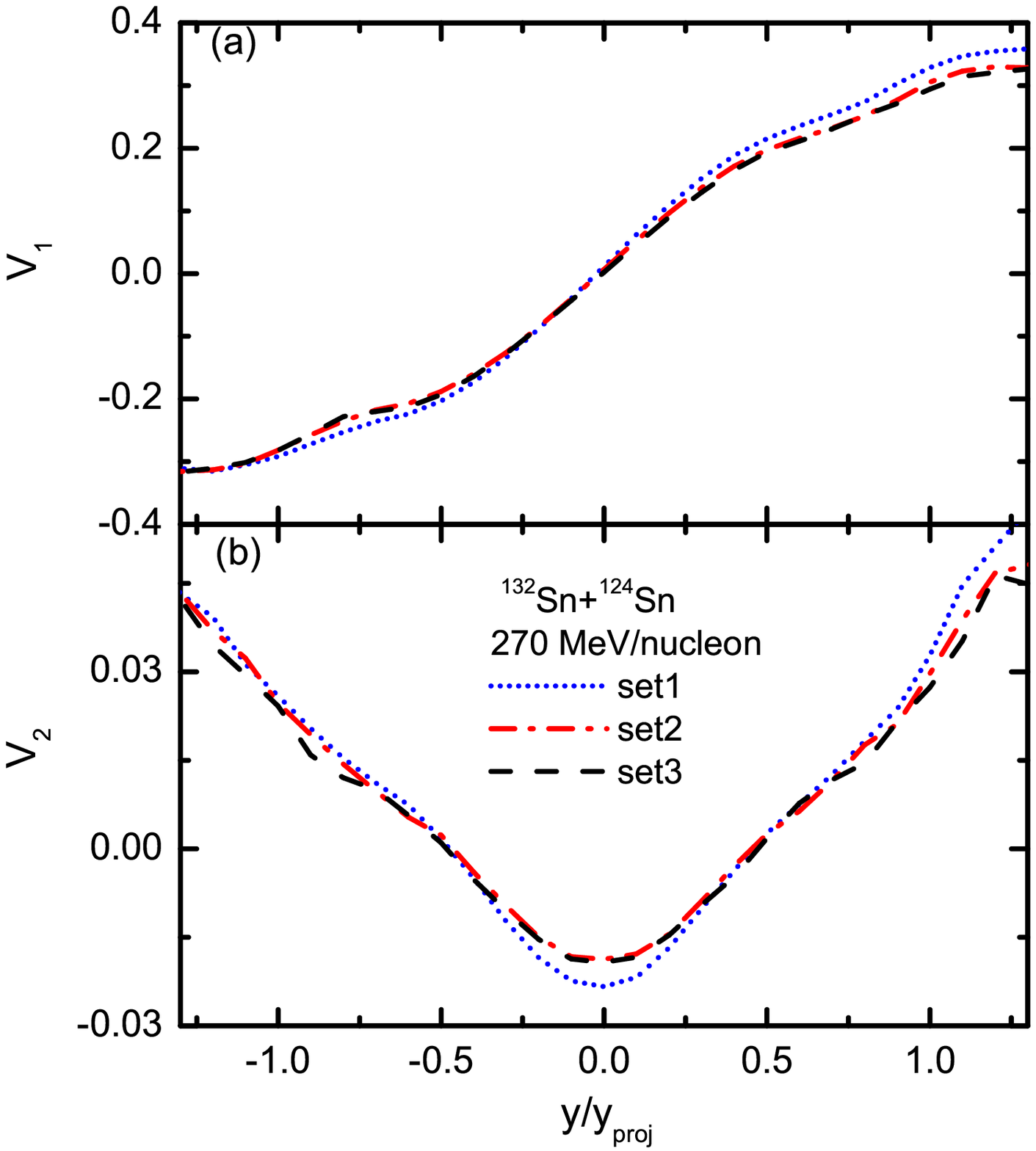}
\caption{(Color online) Rapidity distribution of the collective flows of free neutrons   in the
$^{132}\mathrm{Sn} + ^{124}\mathrm{Sn}$ reaction at an incident energy of 270 MeV/nucleon for impact parameter b=3 fm.}\label{132fl}
\end{figure}
\begin{figure}[H]
\includegraphics[height=10cm,width=8cm]{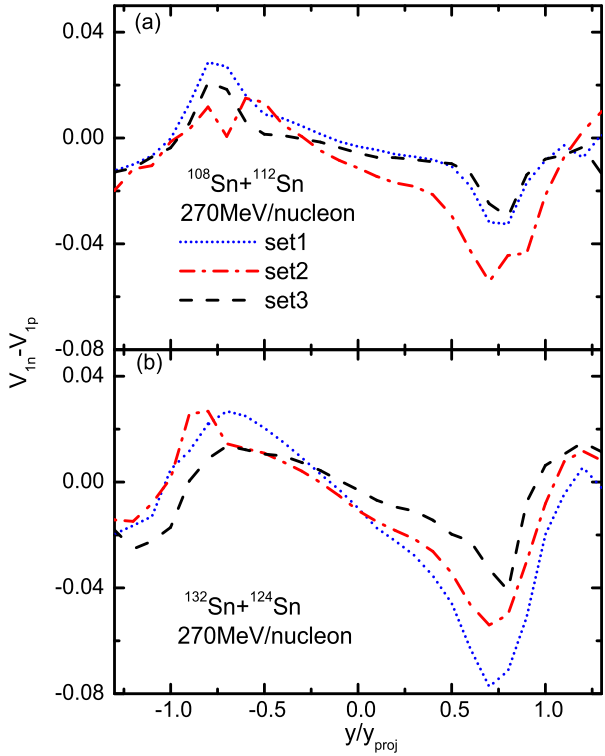}
\caption{(Color online) The difference between neutron and proton directed flows in the
$^{108}\mathrm{Sn} + ^{112}\mathrm{Sn}$ and $^{132}\mathrm{Sn} + ^{124}\mathrm{Sn}$ reactions at an incident energy of 270 MeV/nucleon. }\label{flowm}
\end{figure}
\begin{figure}[H]
\includegraphics[height=6cm,width=8cm]{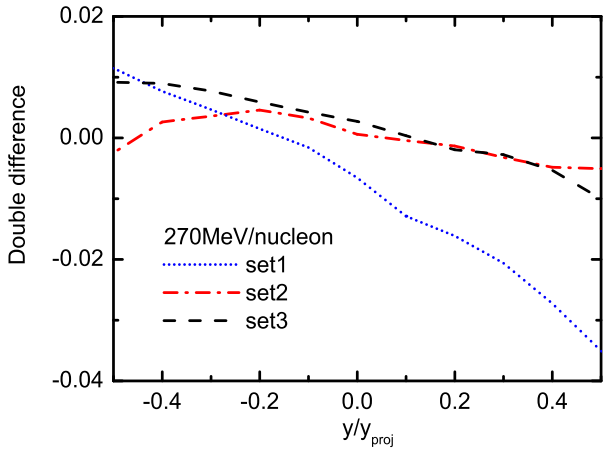}
\caption{(Color online) The double difference of directed flows
 $\rm{(V_{1n}-V_{1p})_{^{132}\mathrm{Sn} + ^{124}\mathrm{Sn}}-(V_{1n}-V_{1p})_{^{108}\mathrm{Sn} + ^{112}\mathrm{Sn}}}$ at an incident energy of 270 MeV/nucleon. }\label{flowm1}
\end{figure}
The $\rm^{108}Sn + ^{112}Sn$ and $\rm^{132}Sn + ^{124}Sn$ collisions of this work are investigated  at  the incident energies of 270 $A$ MeV and  impact parameter b=3 fm. At  incident energies of 270 $A$ MeV, the nuclear matter of  collision central can be compressed to densities approaching 2$\rho_0$. At this dense region, the collective flows, which reflect the repulsion interaction, may  depend on the slope of symmetry energy.  The directed and elliptic flows of $\rm^{108}Sn + ^{112}Sn$ are shown in Fig.\ref{108fl}, and the directed and elliptic flows of $\rm^{132}Sn + ^{124}Sn$ are shown in Fig.\ref{132fl}. It is reasonable that the directed flow $V_1$ is an order of magnitude larger than the  elliptic flow $V_2$. In the same  reaction system, the difference of  directed flows with various slope of symmetry energy (set1, set2 and set3) is small. The difference of elliptic flows with various slope of symmetry energy is also small. In order to find the relationship between the slope of symmetric energy and the collective flow, we need to do some processing on the data of collective flow.

It has been realized that the difference between the neutron and proton directed flows emitted from heavy-ion collisions can be used to extract the density dependence of symmetry energy. The difference between the neutron and proton directed flows is defined as $\rm{V_{1n}-V_{1p}}$.  The difference between the neutron and proton directed flows of  $\rm^{108}Sn + ^{112}Sn$ and $\rm^{132}Sn + ^{124}Sn$ collisions is shown in Fig.\ref{flowm}.  It is worth mentioning that the trend and shape of the difference between the neutron and proton directed flows is similar to previous study\cite{rmq52}.  For a given single reaction system (nearly symmetric $^{108}\mathrm{Sn} + ^{112}\mathrm{Sn}$ system or neutron rich $^{132}\mathrm{Sn} + ^{124}\mathrm{Sn}$ system), the difference between the neutron and proton directed flows seems to be disorganized.  When the difference between the neutron and proton directed  flows of the nearly symmetric and neutron rich systems are compared side by side,   the difference between soft symmetry energy and stiff symmetry energy in the neutron-rich system is inverted slightly from the more symmetric system. Therefore, we define
a double difference of directed flows as  $\rm{(V_{1n}-V_{1p})_{^{132}\mathrm{Sn} + ^{124}\mathrm{Sn}}-(V_{1n}-V_{1p})_{^{108}\mathrm{Sn} + ^{112}\mathrm{Sn}}}$. As shown in Fig.\ref{flowm1}, the double difference of directed flows increases with increasing the slope of symmetry energy in the midrapidity region.

\begin{figure}
\includegraphics[height=8cm,width=8cm]{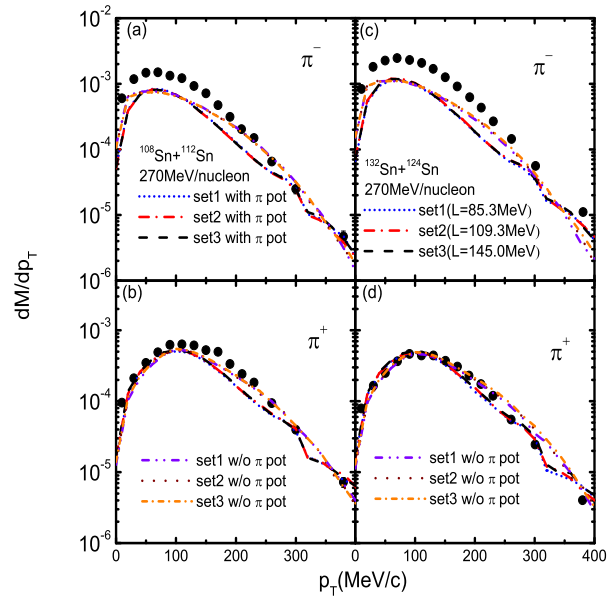}
\caption{(Color online) The transverse momentum spectra of pion as a function of transverse momentum at  an incident energy of 270 MeV/nucleon.  The left two panels [(a) and (b)] are the results of  $^{108}\mathrm{Sn} + ^{112}\mathrm{Sn}$  reaction, and the right two panels [(c) and (d)] are the results of  $^{132}\mathrm{Sn} + ^{124}\mathrm{Sn}$ reaction.}\label{mpi}
\end{figure}

Apart form the collective flows,  the  production of isospin exotic  particles, such as hyperons, kaons and pions, can also be used to extract the symmetry energy\cite{rmq17,rmq18,rmq19,rmq20,rmq21,rmq22,rmq23}.  Since incident energy of 270 $A$ MeV is much smaller than the threshold energy of hyperons and  kaons, the isospin exotic particles are mainly pions. In this work, we calculate the properties of pion in two cases. One is to calculate properties of pion with $\pi$ potential, and the other is to calculate properties of pion  without $\pi$ potential.
As shown in Fig.\ref{mpi},  the left and right panels are the transverse momentum spectra of pion for the nearly symmetric $^{108}\mathrm{Sn} + ^{112}\mathrm{Sn}$ and the neutron rich  $^{132}\mathrm{Sn} + ^{124}\mathrm{Sn}$ reactions at $\theta_{cm}<90^{\circ}$, respectively. For collisions between isotopes, the $\pi^+$ is mainly generated from the collisions between protons, and the $\pi^-$ is mainly generated from the collisions between neutrons. Since the number of protons is the same for isotopes, there will be no significant difference in the yield of $\pi^+$. As shown in lower panels (b) and (d) of Fig.\ref{mpi}, the transverse momentum spectra of $\pi^+$ is close to each other for various systems and slopes of symmetry energy. However, since  there are more neutron-neutron scatterings in  neutron-rich system, there is a difference between the nearly symmetric $^{108}\mathrm{Sn} + ^{112}\mathrm{Sn}$ system and the neutron rich $^{132}\mathrm{Sn} + ^{124}\mathrm{Sn}$ system.
As shown in  upper panels (a) and (c) of Fig.\ref{mpi}, the transverse momentum spectra of $\pi^-$ in the neutron rich $^{132}\mathrm{Sn} + ^{124}\mathrm{Sn}$ system is higher than one in the nearly symmetric $^{108}\mathrm{Sn} + ^{112}\mathrm{Sn}$ system. Besides, theoretically,  a stiffer symmetry would have a stronger repulsive force to push out neutrons resulting in decreasing the  $\pi^-$ yield. However, in this work, the transverse momentum spectra of $\pi^-$ is not sensitive to the slope of symmetry energy. This result may be due to the fact that the symmetry energy of various slope parameter does not differ greatly at densities less than 2$\rho_0$.
Moreover,   $\pi^+$ is consistent with the S$\pi$RIT data\cite{rmq34}, however,  the transverse momentum spectra  $\pi^-$ is lower than the S$\pi$RIT data. As shown in Fig.\ref{mpi}, the effect of $\pi$ potential on the transverse momentum spectra of $\pi$ is obvious. The $\pi$ potential potential in the medium, which is not well understood up to now, may be the reason why the predictions of  $\pi^-$ is lower
than the experiment data.

\begin{figure}
\includegraphics[height=10cm,width=8cm]{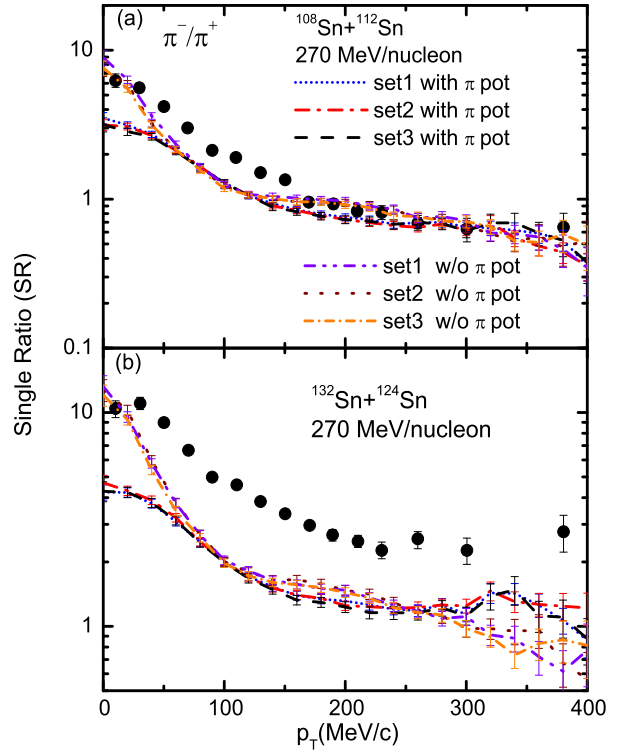}
\caption{(Color online) Single spectral ratios of  pion as a function of transverse momentum for the
$^{108}\mathrm{Sn} + ^{112}\mathrm{Sn}$ and $^{132}\mathrm{Sn} + ^{124}\mathrm{Sn}$ reactions at  an incident energy of 270 MeV/nucleon.}\label{sratio}
\end{figure}
For a given symmetry energy,  since the symmetry potential energy is repulsive for neutrons and $\pi^-$ and attractive for protons and $\pi^+$ in the neutron-rich matter, the  single ratio SR($\pi^-/\pi^+$)=$[dM(\pi^-)/dp_T]/[dM(\pi^+)/dp_T]$ may obviously differ for the nearly  and the neutron rich systems.  As shown in  Fig.\ref{sratio}, the  single ratio of the neutron rich $^{132}\mathrm{Sn} + ^{124}\mathrm{Sn}$ system is higher than one of the nearly symmetric $^{108}\mathrm{Sn} + ^{112}\mathrm{Sn}$ system. However, the single ratio of this work is not sensitive to the slope of symmetry energy for a given nearly symmetric system or neutron rich system. This is mainly originated from the fact that the difference of symmetry energy with set1, set2 and set3 is not obvious  at density below $2\rho_0$.
Moreover, the  single ratio of $^{108}\mathrm{Sn} + ^{112}\mathrm{Sn}$ is lower than but not far away from the experiment data at $p_T<200$ $\mathrm{MeV}/c$,   and is consistent with the experiment data at $p_T>200$ $\mathrm{MeV}/c$.  However, the  single ratio of $^{132}\mathrm{Sn} + ^{124}\mathrm{Sn}$ is  lower than the experimental data at the entire $p_T$ domain. This is due to the fact that the transverse momentum spectra $\pi^-$ of $^{132}\mathrm{Sn} + ^{124}\mathrm{Sn}$ is lower than the experiment data.

\begin{figure}
\includegraphics[height=6cm,width=8cm]{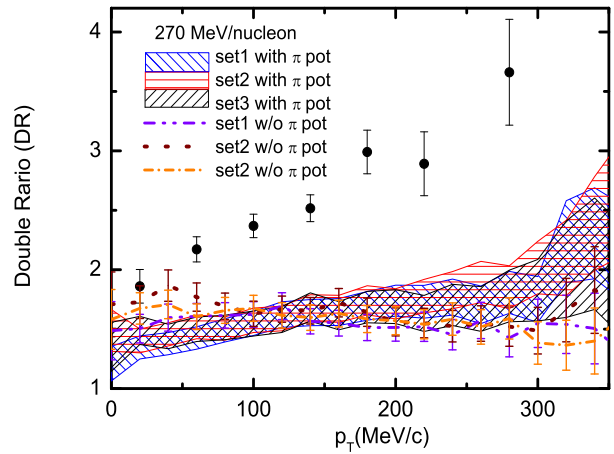}
\caption{(Color online)  The double ratio of pion as a function of transverse momentum at  an incident energy of 270 MeV/nucleon.}\label{dratio}
\end{figure}

The double ratio between the neutron rich  system and the nearly symmetric system $\mathrm{DR}(\pi^-/\pi^+)=\mathrm{SR}(\pi^-/\pi^+)_{132+124}/\mathrm{SR}(\pi^-/\pi^+)_{108+112}$ , which can  cancel out most of the systematic errors caused by  Coulomb and isoscalar interactions, is thought to be sensitive to the properties of  the symmetry  energy. However, as shown in Fig.\ref{dratio}, with considering the $\pi$ potential, the  double ratios of various slope parameters (set1, set2 and set3) are overlapping with each other.
This is due to the fact that  the symmetry energy of set1, set2 and set3 is similar to each  at density below $2\rho_0$.   The double ratio without the $\pi$ potential decreases with increasing the  transverse momentum, however,  the double ratio with the $\pi$ potential increases with increasing the  transverse momentum. The  increasing trend of double ratio without $\pi$ potential energy is opposite to one of the experimental results, however, the increasing trend of double ratio with $\pi$ potential energy is consistent with  one of the experimental results. Therefore, when the understanding of the $\pi$  potential in the medium becomes clear, the double ratio can be used to extract the properties of symmetry energy.

\section{Conclusions}
The RMF with the different slope parameter of symmetry energy, namely set1, set2 and set3,  has been implemented into the LQMD transport model.  The collective flows of the nearly symmetric $^{108}\mathrm{Sn} + ^{112}\mathrm{Sn}$  and the neutron rich $^{132}\mathrm{Sn} + ^{124}\mathrm{Sn}$ systems have been successfully generated from the LQMD.RMF. It has been observed that the directed flow $V_1$ is an order of magnitude larger than the elliptic flow $V_2$. For a given system, the directed flow $V_1$ and the elliptic flow $V_2$ are close to each other for various slopes of symmetry energy. To explore the relationship between the collective flow and the slope of the symmetric energy, we have defined a difference between the neutron and proton directed flows $\rm{V_{1n}-V_{1p}}$  and a double difference $\rm{(V_{1n}-V_{1p})_{^{132}\mathrm{Sn} + ^{124}\mathrm{Sn}}-(V_{1n}-V_{1p})_{^{108}\mathrm{Sn} + ^{112}\mathrm{Sn}}}$.  When the difference between the neutron and proton directed  flows of the nearly symmetric and neutron rich systems are compared side by side, it is found that  the double difference of directed flows increases with increasing the slope of symmetry energy in the midrapidity region.

We  also investigate the relationship between the isospin exotic particles and the symmetry energy. At incident energies of 270 $A$ MeV,  since the generation threshold of hyperons and kaons is not reached, the isospin exotic particles are pions in this work. The transverse momentum spectra of $\pi^+$ is consistent with the experiment data, however, the transverse momentum spectra of $\pi^-$ is a little lower than the experiment data. This lower transverse momentum spectra of $\pi^-$ will cause lower single ratio  and lower double ratio than the experiment data.
The lower transverse momentum spectra of $\pi^-$ may be due to the poor understanding of the $\pi$ potential in the medium. For given reaction system, the transverse momentum spectra of $\pi^+$ and $\pi^-$  is almost the same for various slopes of symmetry energy in this work. As a consequence, the single ratio  and the double ratio are also similar to each other. This is due to the fact that the symmetry energy of set1, set2 and set3 is similar to each at density below $2\rho_0$. The effect of $\pi$ potential on the properties of $\pi$ is obvious.
Especially,  the double ratio without the $\pi$ potential decreases with increasing the  transverse momentum, however,  the double ratio with the $\pi$ potential increases with increasing the  transverse momentum.
When the understanding of the $\pi$ potential in the medium becomes clear, we believe that the properties of isospin exotic particles can be used to extract the symmetry energy.

\section*{ACKNOWLEDGMENT}
This work was supported by the National Natural Science Foundation of China (Projects Nos 12147106, 12175072 and 11722546) and the Talent Program of South China University of Technology (Projects No. 20210115).

\appendix
\section{DETAILS of EQUATION OF MOTION}
For numerical calculation, the equation of motion (Eq.(\ref{rqmd8}))  needs to be written in  computed form. With Eq.(\ref{rqmd9}) and $M_i^*=M_i-S_i=M_N-S_i$ as inputs, the  equation of motion (Eq.(\ref{rqmd8})) can be  expanded as
\begin{eqnarray}
&\dot{\vec{r}}_i=&\frac{\vec{p}_i^*}{p_i^{*0}}+\sum_{j\neq i}[D_{ij}\frac{\partial\rho_{ij}}{\partial\vec{p}_i}
+D_{ji}\frac{\partial\rho_{ji}}{\partial\vec{p}_i}\nonumber\\&&
+(D_j\frac{\partial f_i}{\partial\vec{p}_i}+A_{ji}^\mu\frac{\partial z_{i\mu}}{\partial\vec{p}_i})\rho_{ji}\nonumber\\&&
+(D_j't_i\frac{\partial f_i}{\partial\vec{p}_i}+A_{ji}'^{\mu}\frac{\partial z_{i\mu}}{\partial\vec{p}_i})\rho_{ji}],
\label{ad1}
\end{eqnarray}

\begin{eqnarray}
\vec{\dot{p}}=-\sum_{j\neq i}[D_{ij}\frac{\partial\rho_{ij}}{\partial\vec{r}_i}
+D_{ji}\frac{\partial\rho_{ji}}{\partial\vec{r}_i}],
\label{ad2}
\end{eqnarray}
with
\begin{eqnarray}
D_{ij}=D_if_j+A_{ij}^{\mu}z_{j\mu}+D_i't_jf_j+A_{ij}'^{\mu}z_{j\mu}
\label{ad3}
\end{eqnarray}
\begin{eqnarray}
D_{i}=-\frac{g_\sigma}{2}\frac{M_i^*}{p_i^{*0}}\frac{\partial \sigma_i}{\partial\rho_{Si}}
\label{ad4}
\end{eqnarray}
\begin{eqnarray}
A_{ij}^\mu=\frac{g_\omega^2}{2m_\omega^2}B_iB_jz_i^{*\mu}
\label{ad5}
\end{eqnarray}
\begin{eqnarray}
D_{i}'=-\frac{g_\delta}{2}t_i\frac{M_i^*}{p_i^{*0}}\frac{\partial \delta_i}{\partial\rho_{S3,i}}
\label{ad6}
\end{eqnarray}
\begin{eqnarray}
A_{ij}'^\mu=\frac{g_\rho^2}{2m_\rho^2}t_it_jB_iB_jz_i^{*\mu}
\label{ad7}
\end{eqnarray}
where $z_i^{\mu}=p_i^{\mu}/p_i^{0}$ and $z_i^{*\mu}=p_i^{*\mu}/p_i^{*0}$. Based on Eq.(\ref{rqmd10}), $\frac{\partial\sigma_i}{\partial\rho_{Si}}$ and $\frac{\partial\delta_i}{\partial\rho_{S3,i}}$ are obtained as follow:
 \begin{eqnarray}
\frac{\partial\sigma_i}{\partial\rho_{Si}}=\frac{g_\sigma}{m_\sigma^2+2g_2\sigma_i+3g_3\sigma_i^2},\quad
\frac{\partial\delta_i}{\partial\rho_{S3,i}}=\frac{g_\delta}{m_\delta^2}
\end{eqnarray}

In  two-body center-of-mass frame,  $\rho_{ij}$ equals $\rho_{ji}$. The distance squared $q_{T,ij}^2$ reduces to
 $q_{T,ij}^2\equiv -\vec{r}_{ij}^2-\frac{(\vec{r}_{ij}\cdot \vec{p}_{ij})^2}{p_{ij}^2}$.
In actual calculation,  we have replaced $p_i^0$ with $\sqrt{\vec{p}_i^2+M_i^2}$ to save calculation time\cite{rmqa6}. In doing so, the partial derivative of density versus momentum and space can be written as
 \begin{eqnarray}
\frac{\partial\rho_{ij}}{\partial \vec{p}_i}&=&-\frac{\rho_{ij}}{2L}\frac{(\vec{r}_{ij}\cdot\vec{ p}_{ij})\cdot \vec{r}_{ij}}{p_{ij}^2}
-\frac{\rho_{ij}}{2L}\frac{(\vec{r}_{ij}\cdot\vec{ p}_{ij})^2}{p_{ij}^4}\{\vec{p}_{ij}-\frac{\vec{p}_i}{p_i^0}p_{ij}^0\}\nonumber\\&&
+\rho_{ij}\frac{\gamma_{ij}^2\vec{\beta}_{ij}}{p_i^0+p_j^0}[1-\vec{\beta}_{ij}\frac{\vec{p}_i}{p_i^0}],
\label{ad7}
\end{eqnarray}
 \begin{eqnarray}
\frac{\partial\rho_{ij}}{\partial \vec{r}_i}
=-\frac{\rho_{ij}}{2L}[\vec{r}_{ij}+\frac{(\vec{r}_{ij}\cdot \vec{p}_{ij})\cdot \vec{p}_{ij}}{p_{ij}^2}],
\label{la14}
\end{eqnarray}
where $\vec{\beta}_{ij}$ is defined as $(\vec{p}_i+\vec{p}_j)/(p_i^0+p_j^0)$. Besides, $\frac{\partial f_i}{\partial \vec{p}_i}$ can be written as $-\frac{M_i\vec{p}_i}{p_0^3}$. The partial derivative of energy part of $\frac{\partial z_{i\mu}}{\partial \vec{p}_i}$ is zero, and the partial derivative of momentum part of $\frac{\partial z_{i\mu}}{\partial \vec{p}_i}$  is written as $\frac{-p_i^2}{p_i^{03}}$. With those above equations, the momentum and space of neutrons and protons are known for sure.

\end{document}